\documentclass{webofc}
\usepackage[varg]{txfonts}   
\newcommand{ \be }{\begin{equation}}
\newcommand{ \ee }{\end{equation}}
\newcommand{ \bea }{\begin{eqnarray}}
\newcommand{ \eea }{\end{eqnarray}}
\newcommand{ \la }{\langle}
\newcommand{ \ra }{\rangle}
\newcommand{ \bs }{{\bf s}}
\newcommand{ \bl }{{\bf l}}
\newcommand{ \bv }{{\bf v}}
\newcommand{ \bp }{{\bf p}}

\newcommand{ \bz }{{\bf z}}
\newcommand{ \bB }{{\bf B}}
\newcommand{ \bJ }{{\bf J}}

\newcommand{ \rT }{{\rm T}}

\newcommand{\mean}[1]{\la #1 \ra}

\newcommand{ \grad }{{\bf \nabla}}

\newcommand{ \bomega }{{\boldsymbol{\omega}}}

\begin{document}

\title{Vorticity and particle polarization in heavy ion collisions
  (experimental perspective)}

%

\author{\firstname{Sergei A.} 
\lastname{Voloshin}\inst{1}\fnsep\thanks{\email{sergei.voloshin@wayne.edu}} 
}

\institute{Wayne State University, 666 W. Hancock, Detroit 48201,
  Michigan, U.S.A.
          }

\abstract{The recent measurements of the global polarization and
  vector meson spin alignment along the system orbital momentum in
  heavy ion collisions are briefly reviewed. A possible connection
  between the global polarization and the chiral anomalous effects is
  discussed along with possible experimental checks. Future
  directions, in particular those aimed on the detailed mapping of the
  vorticity fields, are outlined. The Blast Wave model is used for an
  estimate of the anisotropic flow effect on the vorticity component
  along the beam direction.  We also point to a possibility of a
  circular pattern in the vorticity field in asymmetric, e.g. Cu+Au,
  central collisions. }
\maketitle

\section{Introduction}
\label{intro}
The idea of the global polarization in heavy ion collisions, the
phenomenon characterized by the polarization of the secondary
particles along the global system orbital momentum, is almost 15 years
old. It went ``on-shell'' in 2004~\cite{Liang:2004ph,Voloshin:2004ha}
with the initial predictions for the hyperon polarization as high as
``in the order of tens of a percent''~\cite{Liang:2004ph}. The first
measurements~\cite{ Abelev:2007zk} by the STAR Collaboration of the
lambda hyperon polarization in Au+Au collisions at 200 GeV put an
upper limit on hyperon polarization of about $|P_\Lambda|\le
0.02$. Subsequently, the theoretical predictions have been
improved~\cite{Gao:2007bc}, especially with a better understanding of
the statistical mechanics of vortical fluid with non-zero spin
particles~\cite{Becattini:2007nd}, and development of the
hydrodynamical calculations assuming local angular momentum
equilibrium.  Rough estimate of the polarization can be obtained
with the help of a simple nonrelativistic expression for a particle
distribution in a fluid with nonzero
vorticity~\cite{Becattini:2016gvu} (for a strict relativistic
consideration see~\cite{Becattini}):
\begin{equation}
p\propto \exp\left[-E/T-\bomega (\bs+\bl)/T -\mu B/T\right],
\label{eq:nr}
\end{equation}  
where $\bomega=\frac{1}{2}\grad\times \bv$ is the nonrelativistic
vorticity, and $\bv$ is the fluid velocity. Figure 1 shows a cartoon of
a non-central nuclear collision with arrows indicating the velocity
field of the matter at the plane $z=0$. Just ``guessing'' that
the difference in velocities in the ``upper'' and ``lower'' parts of the
system is about a few tenths of the speed of light and that the
transverse size of the system is about 10~fm, one would conclude that
the vorticity might be at the level of a few percent of $\rm
fm^{-1}$. Then the nonrelativistic formula~(\ref{eq:nr}) yields for
the spin 1/2 particle polarization, $P \approx \omega/(2T)$, in the
range of a few percent (assuming $T\sim 100$~MeV).

\begin{figure}
\centering
\sidecaption
\includegraphics[width=8.5cm]{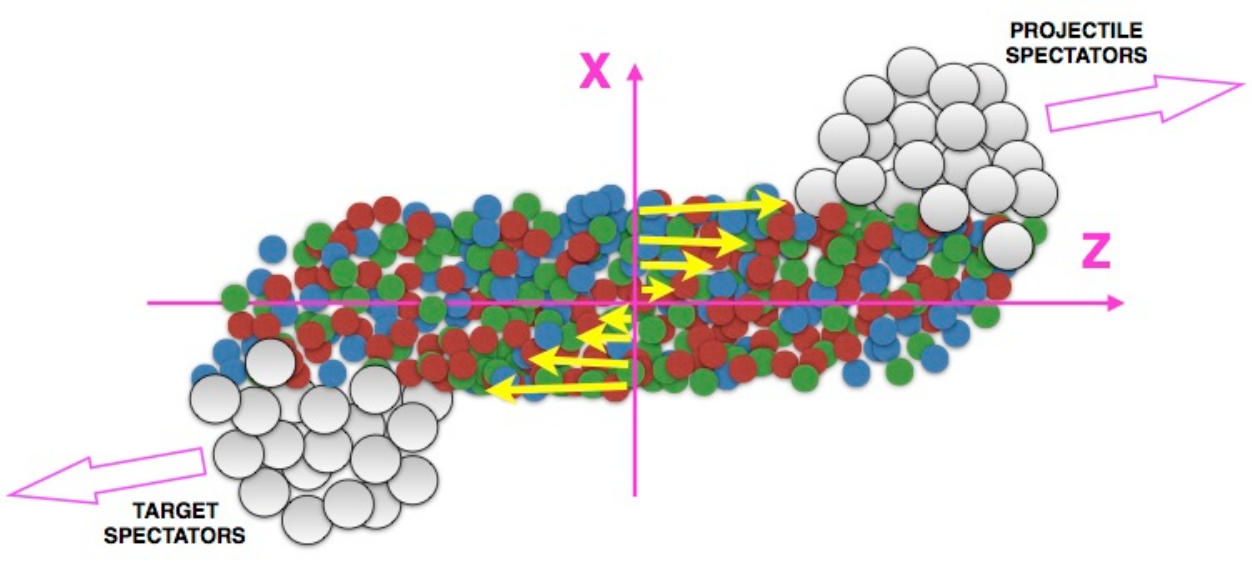}
\hspace*{-0.5cm}
\caption{Cartoon of a non-central nuclear collision. The arrows indicate
the collective velocity of the matter at $z=0$ plane.}
\label{fig:collision} 
\end{figure}

\begin{figure}
\centering
\sidecaption
\hspace*{1cm}
\includegraphics[width=5.cm,angle=0]{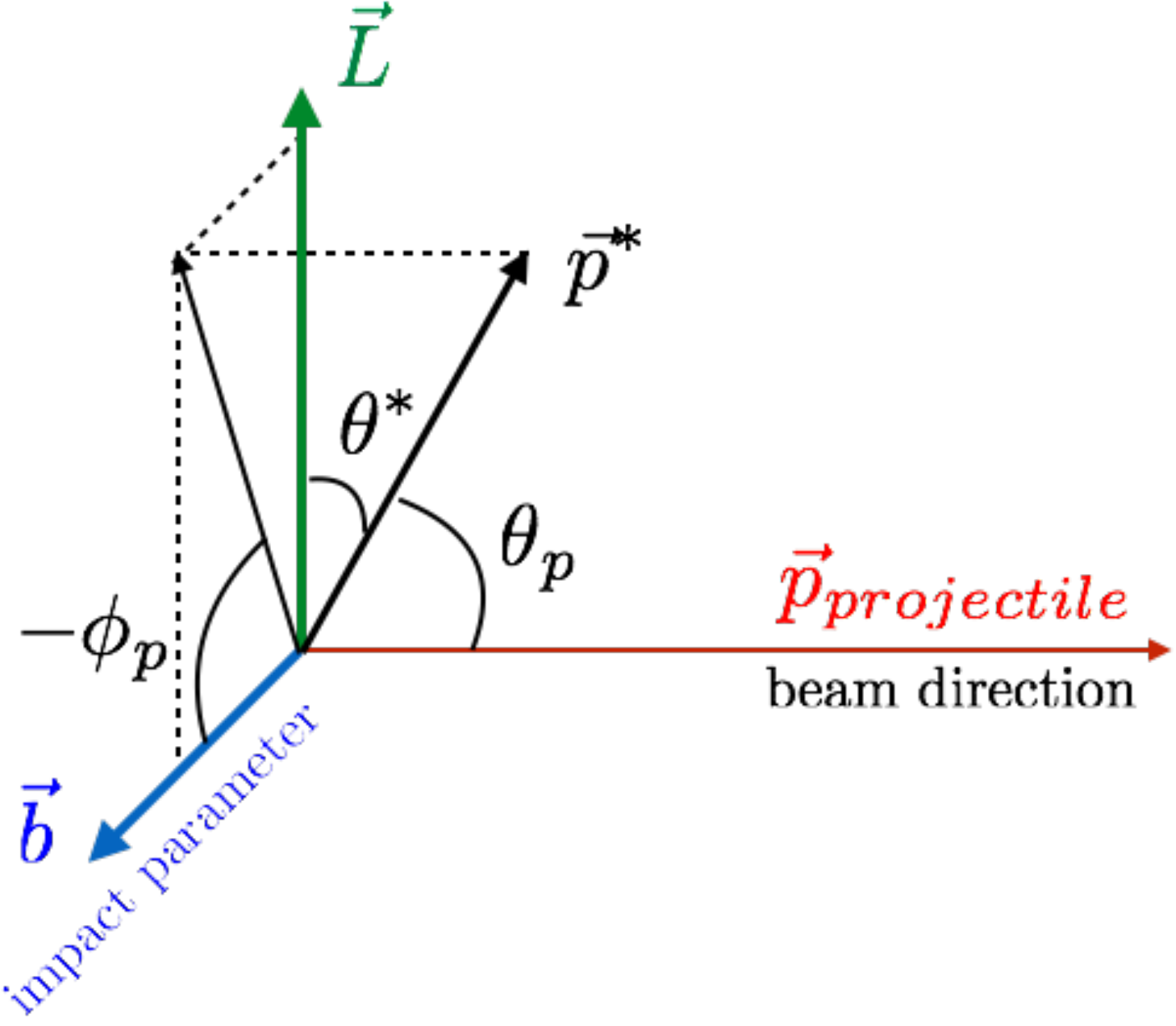}
\caption{Diagram explaining the notations for different angles
  discussed in the text. $\bp^*$ is the proton momentum in the hyperon
  rest frame. Vertical direction is the direction of the global
  orbital momentum -- the global polarization direction. $\phi_p$ is
  the proton emission azimuth in the system with $x-z$ plane aligned
  with the reaction plane.}
\label{fig:coordinates}       
\end{figure}

The simplest way to measure the global polarization is via analysis of
the angular distributions of the products of weakly decaying
hyperons. Weak interaction violates parity, and, e.g. in the lambda
hyperon decay the protons are emitted preferentially in the direction
of the lambda's spin:
\be 
\frac{dN}{d\cos\theta^*} \propto 1+\alpha_H P_H\cos\theta^* ,
\ee
where $\theta^*$ is the polar angle of the proton emission relative to
the polarization direction in the hyperon rest frame, $-1\le P_H \le
1$ is the hyperon polarization, and the parameter
$\alpha_\Lambda=-\alpha_{\bar{\Lambda}}\approx0.624$.  To measure the
polarization of strongly decaying particles is obviously significantly
more difficult. It is not at all possible for spin 1/2 particles; for
the vector mesons one can hope to measure the deviation from 1/3 of
the probability for the spin projection to be
zero~\cite{Liang:2004xn}. The angular distribution (averaged over the
azimuthal distribution around the polarization direction) of the decay
products in this case reads:
\begin{eqnarray}
\frac{dN}{d\cos\theta^*} &\propto& \rho_{0,0}|Y_{1,0}|^2+ \rho_{1,1}|Y_{1,-1}|^2+
\rho_{-1,-1}|Y_{1,1}|^2\propto \rho_{0,0}\cos^2\theta^* +\frac{1}{2}(\rho_{1,1}+\rho_{-1,-1})\sin^2\theta^*
\\
&\propto&(1-\rho_{0,0})+(3\,\rho_{0,0}-1)\cos^2\theta^*
\end{eqnarray}
where $\rho_{0,0}$, $\rho_{1,1}$, and $\rho_{-1,-1}$ are the
probabilities for the particle to have spin projection on the
direction of polarization to be zero, $+1$, and $-1$,
respectively. The deviation from the non-polarized state value
$\rho_{0,0}=1/3$ is in this case a second order
effect~\cite{Liang:2004xn}. For example, an estimate based on
Eq.~\ref{eq:nr} yields $\rho_{00}=1/[3+(\omega/T)^2]$.

To measure the hyperon global polarization or vector meson spin
alignment experimentally, one can either analyze directly the
distributions in $\theta^*$, or, in case of the global polarization
direction defined by the one of the flow event planes, analyze the
azimuthal distribution of the decay products (in the resonance rest
frame) relative to that flow plane.  The azimuthal distribution
analysis can be often simpler in terms of applying necessary
corrections for the reaction plane resolution and detector
acceptance. In the case of the azimuthal analysis, the polarization
can be calculated as:
\begin{eqnarray}
P_H&=&-\frac{8}{\pi\alpha_H}\mean{\sin(\phi_P^*-\Psi_{\rm RP})},
\\
\rho_{00}&=&\frac{1}{3}-\frac{8}{3}\mean{\cos[2(\phi_p^* -\Psi_{\rm RP})]}.
\end{eqnarray}
For the global polarization, the analysis has to be performed using
one of the first harmonic event planes, with the reaction plane (and
correspondingly, the angular momentum) direction to be determined by
the deflection direction of the projectile spectators (which on
average deflect outward of the collision~\cite{Voloshin:2016ppr}). For
the spin alignment measurements it is possible to use the second order
event plane (which typically has much better resolution).

\section{Results}
\label{sec:results}

The progress in vector spin alignments measurements was presented at
this conference in talks by the STAR and ALICE
Collaborations~\cite{Tu,Mohanty}. The uncertainties in these
measurements are still relatively large, and the results are rather
inconclusive; below I concentrate on the discussion of the global
polarization results.

\begin{figure}[h]
\centering
\includegraphics[width=10cm]{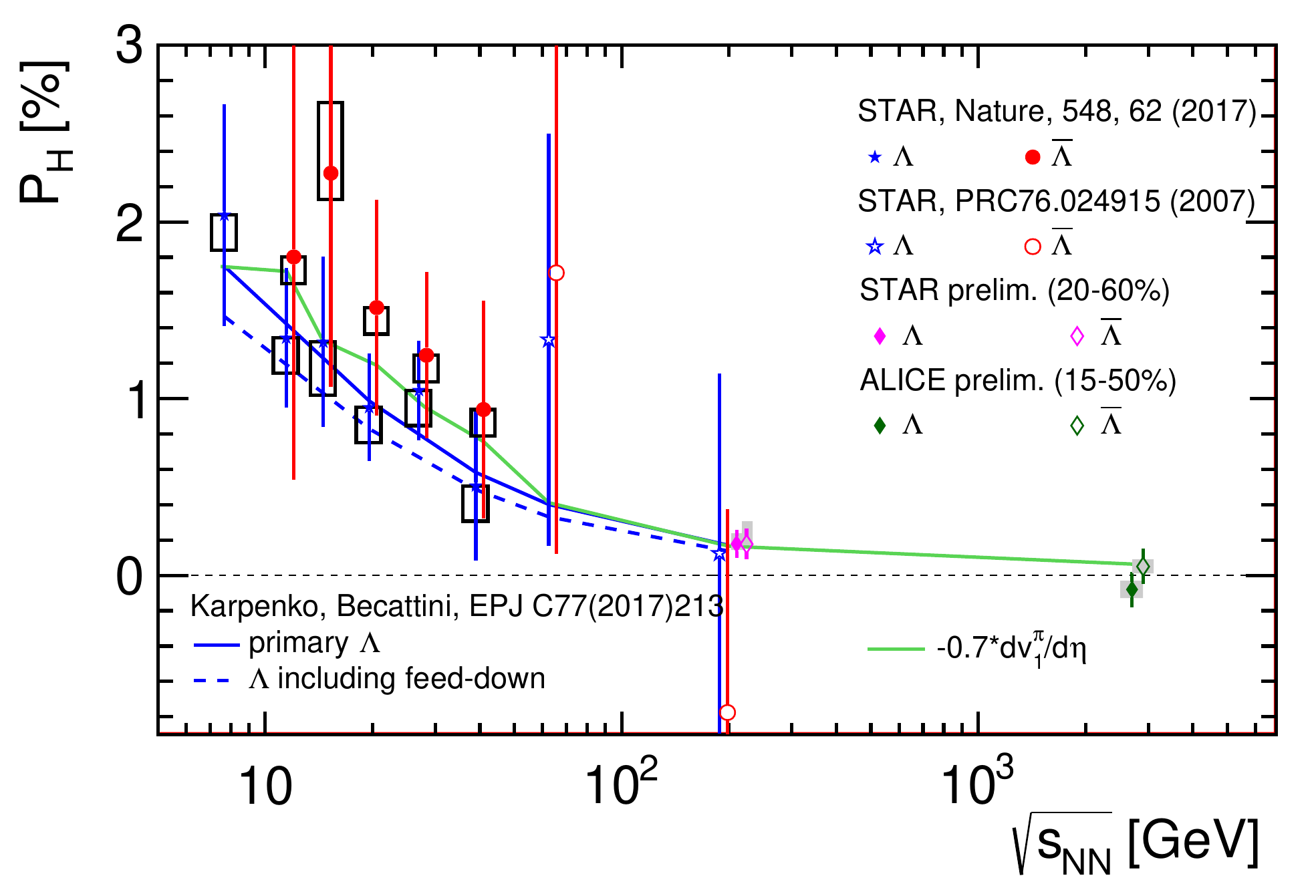}
\caption{Average global polarization of lambda hyperons as a function
  of collision energy. Boxes indicate the systematic uncertainties.}
\label{fig:gpol}  
\end{figure}

Figure~\ref{fig:gpol} shows a compilation of
published\cite{Abelev:2007zk,STAR:2017ckg} and presented at this
conference~\cite{Tu,Mohanty} results on the average global
polarization of lambda and anti-lambda hyperons at mid-rapidity in
mid-central collisions as a function of collision energy.  The blue
solid and dashed lines are the results of hydrodynamic
calculation~\cite{Becattini,Karpenko}, with and without accounting for
the hyperon feed-down contribution, respectively. The procedure for the
feed-down correction is outlined in~\cite{Becattini:2016gvu} with
Eq.~\ref{eq:nr} used for an estimate of the polarization of the higher
spin resonances. Note that the effect of the feed-down correction is
rather modest -- at the level of ~$\sim 15$\%.

At RHIC beam energy scan collision energies the global polarization of
lambdas and antilambdas seems to be slightly
different. Calculations~\cite{Fang:2016vpj} show that the effect of
nonzero baryon chemical potential could result in a split of the
polarization values for particle and antiparticle, but numerically
(assuming the equilibrium in the hadronic gas) this effect expected to
be small.  More realistic explanation for the difference would be the
effect of the magnetic field that is strongly aligned with the
direction of the orbital momentum. Then the global polarization
analysis might become a very sensitive tool for the measurement of the
magnetic field in heavy ion collisions. The present uncertainties in
polarization measurements allow to put only an upper limit on the
magnitude of the magnetic field, $eB \lesssim 0.01 m_\pi^2$, but even
that is of great interest, as at present, the uncertainty in the
magnetic field calculations spans several orders of
magnitude~\cite{McLerran:2013hla}.

There exist no hydrodynamic calculations for the global polarization
at the LHC energies. For that we can develop an empirical prediction
based on the observation that the slope of the directed flow at
mid-rapidity is likely strongly correlated with the vorticity. Indeed,
the calculation~\cite{Becattini:2015ska} support such a
hypothesis. Then one can use the available data on $v_1$
slope~\cite{Singha:2016mna,Abelev:2013cva} (we used that of charged
pions) for an estimate of the global polarization at higher
energies. This is shown by a green line in Fig.~\ref{fig:gpol}. The
ALICE global polarization
measurements~\cite{Mohanty,Konyushikhin2017}, which within the error-bars
at present are consistent with zero, do not contradict such an
estimate. It will be required to improve the uncertainties about factor
of 3--5 to be able to measure the polarization at the level of
an estimate from the directed flow slopes.

\begin{figure*}[h]
\centering
\vspace*{-1.5cm}
\includegraphics[width=6cm,angle=0]{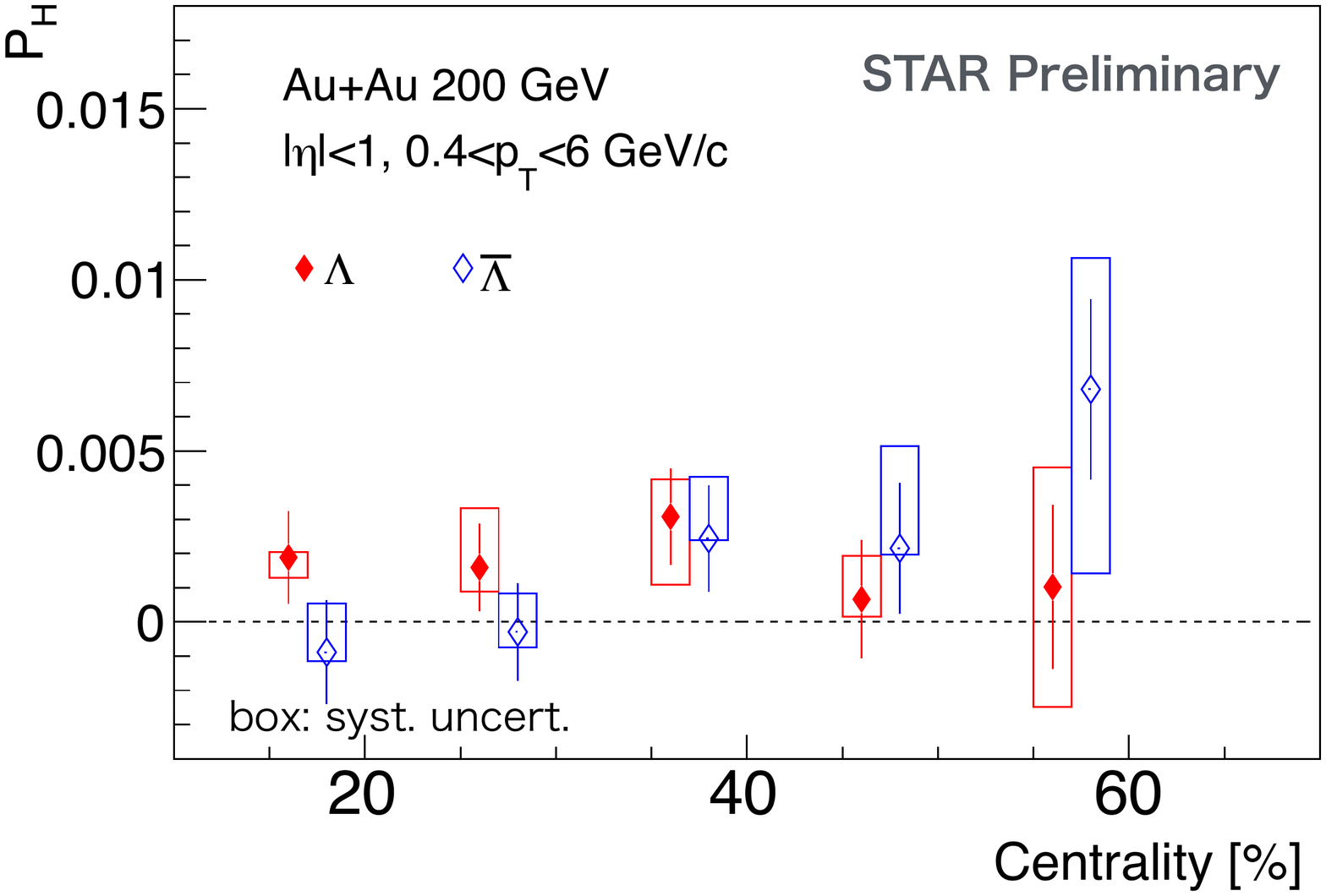}
\includegraphics[width=6cm,angle=0]{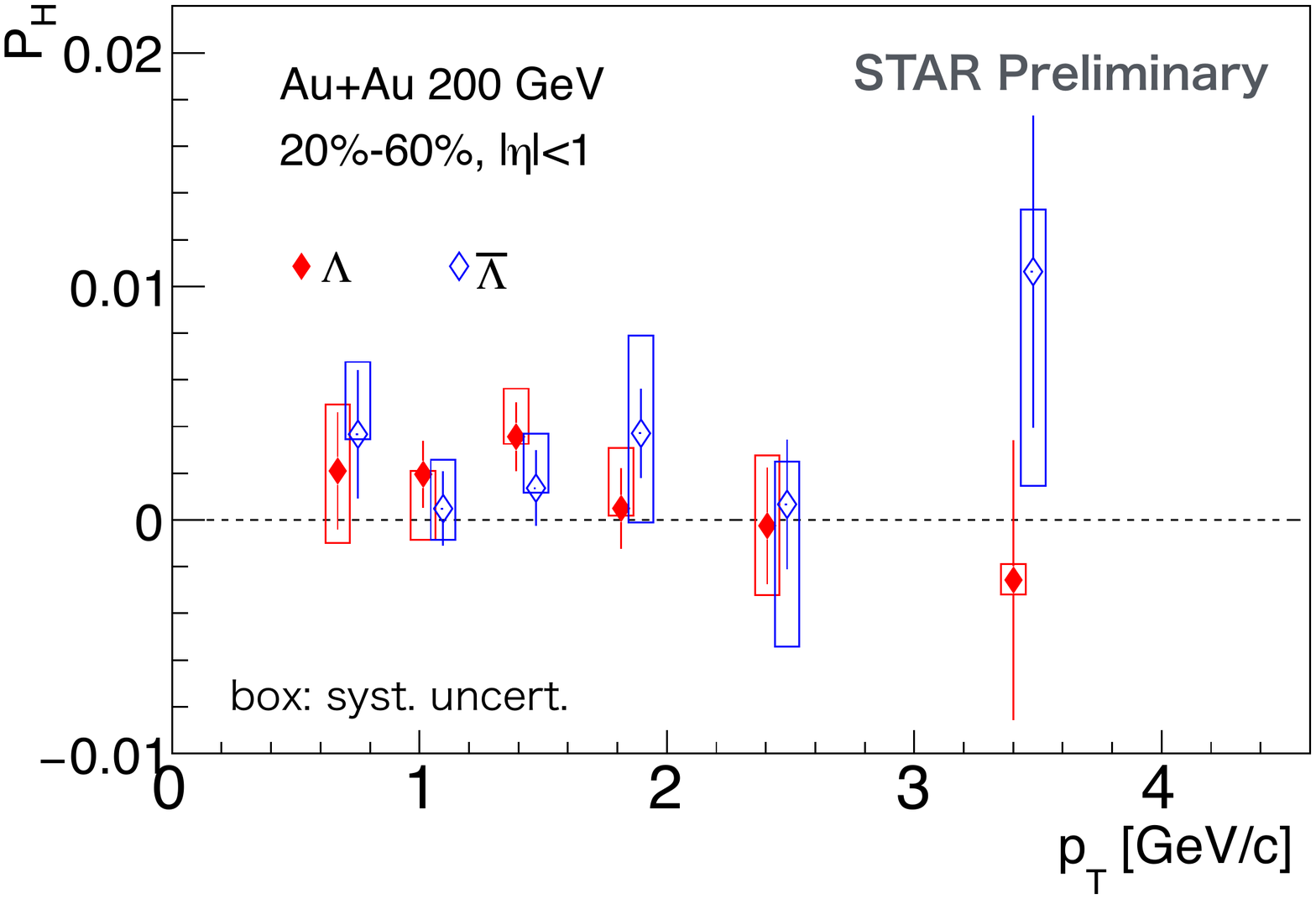}
\vspace*{-2cm}
\caption{Centrality and $p_T$ dependence of the global polarization in
Au+Au collisions at 200~GeV~\cite{Tu,Niida2017}.}
\label{fig:cent_pt}     
\end{figure*}

Global polarization dependence on centrality, the hyperon transverse
momentum, and the orientation of the hyperon emission relative to the
reaction plane have been also of great interest as different models
predict significantly different dependencies. Within the available
statistic, both, the centrality and $p_{\rm T}$ dependencies, see the
preliminary STAR results~\cite{Tu,Niida2017} presented in
Fig.~\ref{fig:cent_pt}, are rather modest. The azimuthal angle
dependence is discussed in more detail in~\cite{Tu}. The preliminary
results indicate stronger polarization of the hyperons emitted
in-plane than those emitted out-of-plane, opposite to the expectations
from the hydrodynamical calculations~\cite{Becattini:2013vja}.

\subsection{Global polarization and chiral anomalous effects}

The exact mechanism relating the vorticity of the medium to the global
polarization is not yet firmly established. It was argued
in~\cite{Baznat:2015eca} that the global polarization appears as a
results of the axial anomalous current:
\be 
\bJ _5 = \left(
\frac{\mu^2+\mu_5^2}{4\pi^2} + \frac{T^2}{12} \right) \bomega 
\ee 
For a more detailed discussion of this possibility, see~\cite{Teryaev} and
references therein.

\begin{figure}[h]
\centering
\vspace{-1.cm}
\includegraphics[width=8cm,angle=90]{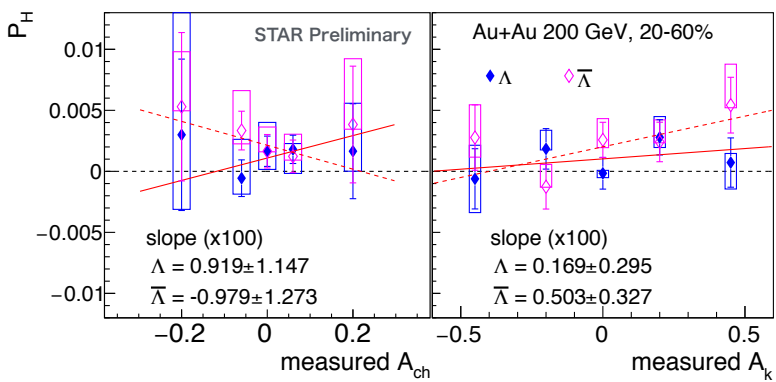}
\vspace*{-1.5cm}
\caption{Global polarization as a function of the event charge
  asymmetry $A_{ch}=(N_+-N_-)/(N_++N_-)$, and the event net kaon ratio
  $A_K=(N_{K^+}-N_{K^-})/(N_{K_+}+N_{K^-})$~\cite{Niida2017}.  }
\label{fig:anom}     
\end{figure}

Axial current can be also ``generated'' by the magnetic field in the
system with nonzero vector charge chemical potential (for a review of
chiral anomalous effects in heavy ion collisions
see~\cite{Kharzeev:2015znc}):
\be
\bJ _5 = \frac{1}{2\pi^2} \mu_V(Qe) \bB,
\ee
which in particular relates the direction of the axial current to the
sign of the chemical potential.  This provides an interesting
possibility for the experimental check of the existence of such a
relationship~\cite{Schlihting} by measuring the global polarization as
a function of the event charge asymmetry or the net kaons, hoping that
those are proportional to either electrical charge or strangeness
chemical potentials. In some sense such a measurement would constitute
a measurement of the first ``half'' of the chiral magnetic wave
mechanism (in which the first step is the separation of the axial
charge and the second step includes the electric currents rearranging
the electric charge and finally leading to the quadrupole charge
configuration, see review~\cite{Kharzeev:2015znc}).  The preliminary
results from the STAR Collaboration~\cite{Niida2017} presented in
Fig.~\ref{fig:anom} are not yet conclusive due to the large
statistical uncertainties.

\subsection{Local vorticity and polarization along the beam direction}

While the system orbital momentum provides an obvious choice for the
average direction of the vorticity (and particle polarization),
locally, the vorticity have in general nonzero all three
components. Those might be generated by a particular process, e.g. jet
propagation~\cite{Betz:2007kg} (not yet studied experimentally) or
simply by the expansion of the initially anisotropic medium, see, for
example, the results in~\cite{Becattini:2015ska} (including those
presented only in the arXiv version of this publication), as well as
calculations in~\cite{Baznat:2015eca,Fang:2016vpj}.  In particular,
the vorticity $z$-component, along the beam direction, might reach
large values~\cite{Becattini:2015ska,Becattini,Karpenko}.

\begin{figure}[h]
\centering
\sidecaption
\includegraphics[width=4.5cm,angle=0]{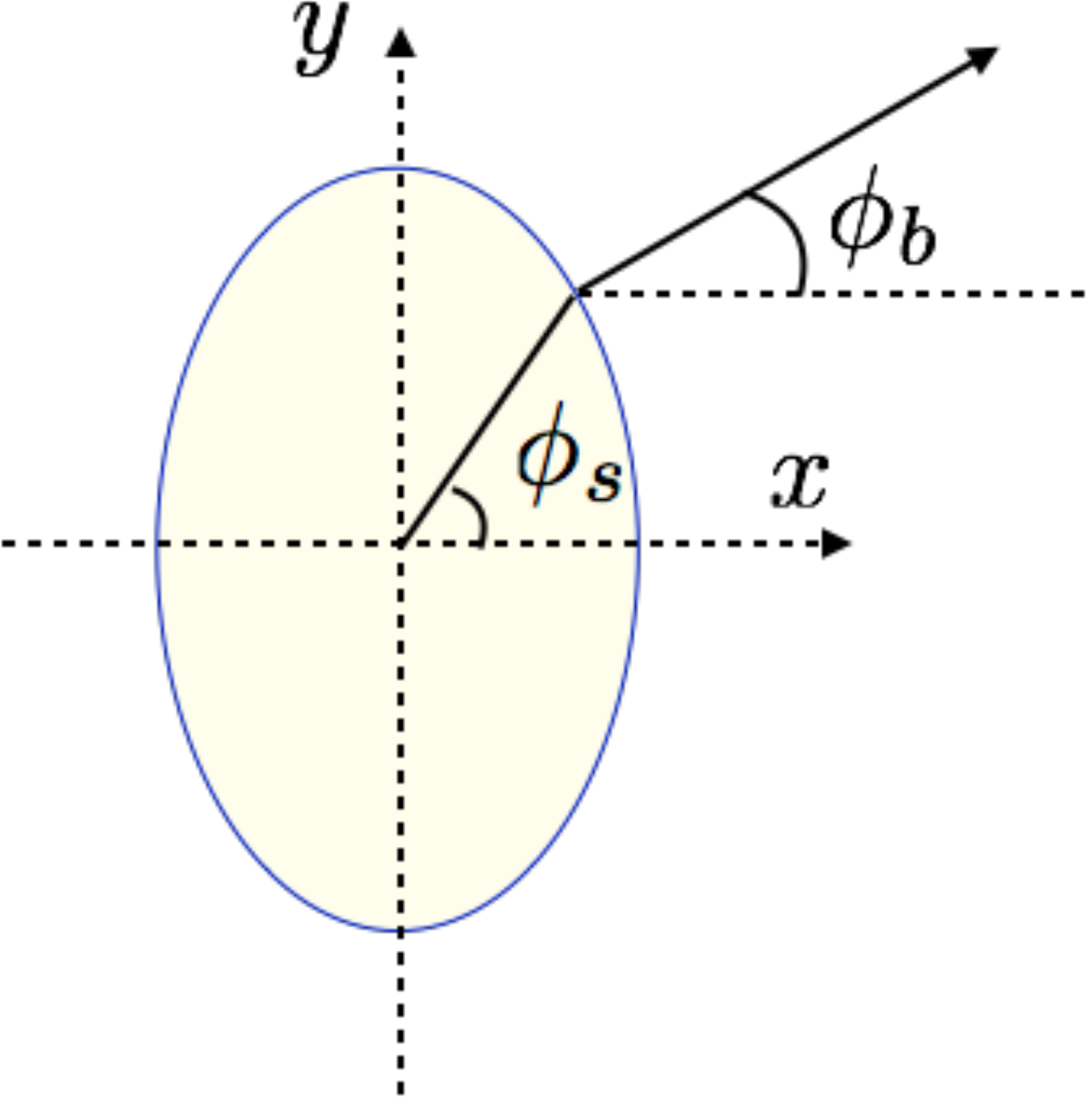}
\caption{The collective velocity of the source element at angle
  $\phi_s$ at the surface is along the boost angle $\phi_b$,
  perpendicular to the surface described by Eq.~\ref{eq:rmax}. The
  boost velocity is given by ~Eq.~\ref{eq:boost}.
}
\label{fig:BW}       
\end{figure}

Below I discuss a simple Blast Wave (BW) model estimates for the
vorticity $z$ component. In a simplest version of BW model that
includes anisotropic flow, the particle production source at
freeze-out is parameterized with 4 parameters: temperature $T$,
maximum radial flow velocity (rapidity), notated as $\rho_0$,
amplitude of azimuthal modulation in expansion velocity, noted below
as $b$, and the spatial anisotropy parameter $a$. The source (see
Fig.~\ref{fig:BW}) is then described by the following equations:
\be
r_{max}=R[1-a \cos(2\phi_s)],
\label{eq:rmax}
\ee
\be
\rho_t=\rho_{t,max}[r/r_{max}(\phi_s)][1+b\cos(2\phi_s)]\approx
\rho_{t,max}(r/R)[1+(a+b)\cos(2\phi_s)].
\label{eq:boost}
\ee
It is assumed that the collective velocity of the source element
located at azimuthal angle $\phi_s$ is boosted with velocity $\rho_t$
perpendicular to the surface of the ellipse similar to that of
Eq.~\ref{eq:rmax}.  Assuming that $a\ll 1,\ b\ll 1$, the difference
$\phi_s-\phi_b \approx 2a\sin(2\phi_s)$ and the vorticity:
\be
\omega_z=1/2 (\grad \times
\bv)_z\approx(\rho_{t,nmax}/R)\sin(n\phi_s)[b_n-a_n].
\label{eq:vortn}
\ee
The estimates above should be valid for anisotropic flow of any
harmonics - which is the reason we have changed in Eq.~\ref{eq:vortn}
the harmonic order from 2 to $n$.  It is obviously quite a rough
approximation (which in principle can be improved) as it leads to a
discontinuity at the origin. It provides the following
estimate for the hyperon polarization:
\be
P_z\approx\omega_z/(2T)\approx 0.1\sin(n\phi_s)[b_n-a_n],
\ee
where we assumed that $\rho_{t,nmax}\sim1$, $R\approx 10$~fm, and
$T\approx 100$~MeV.
In practice, the coefficients $b_n$ and $a_n$ are both of the order of a
few percent, often close to each other. That results in the values for
$z$-polarization not greater than a few per-mill, almost an order of
magnitude lower than obtained in hydrodynamics
calculations~\cite{Becattini,Karpenko}. 

The measurements of the $z$ component of polarization could be
relatively simple as they do not require the knowledge of the first
harmonic event plane. The acceptance effects should be also readily
accounted for requiring that the $z$ component of the polarization
averaged over all azimuthal angles to be zero.

We note that vorticity fields due the anisotropic flow are formed
closer to the freeze-out, unlike the ones due to the ``shear'' in the
initial velocity fields (as shown in ~Fig~\ref{fig:collision}). Having
in mind the finite relaxation time for establishing the equilibrium
the relation between these two vorticities and the final polarizations
can be different.

Finally, we mention another very interesting possibility for vorticity
studies in asymmetric nuclear collisions such as Cu+Au. For relatively
central collisions, when during the collision a smaller nucleus is
fully ``absorbed'' by the larger one (e.g. such collisions can be
selected by requiring no signal in the zero degree calorimeter in the
lighter nucleus beam direction), one can easily imagine a
configuration with toroidal velocity field, and as a consequence, a
vorticity field in the form of a circle. The direction of the
polarization in such a case would be given by
$\hat{\bp_\rT}\times\hat{\bz}$, where $\hat{\bp_\rT}$ and $\hat{\bz}$
are the unit vectors along the particle transverse momentum and the
(lighter nucleus) beam direction.

\section{Summary}

The hyperon polarization measurements provide a unique possibility to
study the velocity fields in heavy ion collisions. The difference in
particle//antiparticle polarization might provide a tool for the
magnetic field measurements. Studing the global polarization as a
function of event net electrical charge or strangeness could establish
the relation between the global polarization and the chiral anomalous
effect.  The future 28 GeV Au+Au RHIC run, as well as the future LHC
runs, will bring a high statistics data sets suitable for many
differential studies outlined in this presentation.

\vspace*{2mm} Very fruitful discussions with F.~Becattini,
Yu.~Karpenko, M.~Konyushikhin, M.~Lisa, T.~Niida, I.~Upsal, and
X.-N.~Wang are greatly appreciated.  This material is based upon work
supported in part by the U.S. Department of Energy Office of Science,
Office of Nuclear Physics under Award Number DE-FG02-92ER-40713.

\end{document}